\documentclass[12pt]{article}
\usepackage[T2A]{fontenc}
\usepackage[koi8-u]{inputenc}
\usepackage[ukrainian,english]{babel}
\input epsf.tex
\newcommand{\nee}[2]{
\begin{equation}
#2
\label{#1} 
\end{equation}
}

\begin{document}
\title{Spherical collapse and mass function of rich clusters 
in models with curvature and cosmological constant}
\author{ Yu. Kulinich, B. Novosyadlyj}
\maketitle

\centerline {Astronomical Observatory of Ivan Franko National University of L'viv}
\centerline {8, Kyryla i Mephodija St., L'viv  79005, Ukraine}

\begin{abstract}
We have analyzed the dependences of the threshold value of amplitude of
linear density fluctuation collapsed at the current epoch, $\delta_c$, 
and its overdensity after virialization, $\Delta_c$, on matter density 
content, 3D curvature parameter and cosmological constant. 
It was shown that in models with negative or zero curvature ($k\le0$) and 
positive cosmological constant ($\Lambda>0$) despite smaller rate of
perturbation growth, $\delta_c$ appears to be smaller than its canonical 
value 1.686 ($k=\Lambda=0$) due to the longer duration of amplitude increase. 
However, in models with positive 3D 
curvature, $\delta_c$ can be larger when the matter density parameter is small. 
In this case critical amplitude
$\delta_c$ approaches infinity when $\Omega_m$ converges to its
small critical value.
Though the range for the threshold values of 
perturbation amplitude is quite narrow - $1.55\le \delta_c\le 1.75$ in the region of 
parameters $-0.4\le\Omega_k\le0.4$, $0\le\Omega_{\Lambda}\le 1$, 
$0.1\le\Omega_m\le 1$, the difference in the concentrations of rich clusters of galaxies 
(as calculated within the Press-Schechter formalism framework) with real and with
canonical values of $\delta_c$ reaches, and for some models exceeds, 30\%. 
The range of changes 
for the overdensity
after virialization, $\Delta_c$, is considerably wider for the same region 
of parameters: $60\le\Delta_c\le180$. 
It results into difference up to $\sim40$\% between
the X-ray temperatures of gas, as calculated for these values and for 
the canonical value of $\Delta_c=178$. Also we have found analytical approximations 
of the dependences $\delta_c(\Omega_m,\Omega_{\Lambda})$ and  
$\Delta_c(\Omega_m,\Omega_{\Lambda})$ with their accuracies for above
mentioned region of parameters being no worse than 0.2\% and 1.7\% respectively.



\end{abstract}

\section{INTRODUCTION}
The mass function and X-ray temperature function of rich clusters of galaxies are
important characteristics of the large-scale structure of the Universe.
The studies of rich clusters are so intensive due to their eminent 
properties: а) they are the largest gravitationally 
bound systems in the Universe; b) the observational 
selection criterion for these structures is quite evident enabling one 
to draw the catalogues of these objects with the highest completeness, at 
least within the radius of $\sim 300h^{-1}$Мpc, within cones $\sim 30^{o}$ around 
the poles of our galaxy; c) space and velocity distributions of galaxies in the 
clusters demonstrate their dynamical equilibrium --  hence indicate to the 
possibility the main theorems of dynamics of gravitational
systems to be applied for determination of their characteristics at current state; d)
Press-Schechter formalism and its modifications [1-9]  
connect these  physical parameters with features
of the power spectrum of initial density perturbations.
This power spectrum is sensible both to dynamical parameters of the Universe
(Hubble constant, 3D-space curvature, cosmological constant), which determine
the rate and acceleration of its expansion, and to the density parameters
of different matter components (baryons, photons, neutrinos
and hypothetical particles which compose the hidden mass of the 
Universe) -- that enables one to determine these parameters on the 
basis of the observed data on the large-scale structure of the Universe.

The latest experimental programs dealing with detection of the microwave background
temperature fluctuations on different angular scales (COBE, Boomerang, MAXIMA, DASI
etc.) and more complete catalogues of galaxies and galaxy clusters
(Sloan Digital Sky Survey, 2dF Galaxy Redshift Survey etc.) 
allow us to determine main cosmological parameters and 
accuracy of such determination.
The enhanced experimental accuracy requires more 
precise theoretical description of the observed phenomena.
Press-Schechter formalism describes dependence of the mass 
function of rich clusters of galaxies on the mean matter density of the Universe 
($\bar\rho$), dispersion of density  fluctuations on the scale of clusters
($\sigma_M$) and threshold value of the amplitude of
density fluctuation collapsed at the current epoch ($\delta_c$):
$n_{PS}(M)=F(M,\bar\rho,\sigma_M)\delta_c/\sigma_M\exp(-\delta_c^2/2\sigma_M^2)$,
where $F=\sqrt{2/\pi}\bar\rho/M^2\;|d\;ln\sigma_M/d\;lnM|$. 
Ecke et al. (1996) \cite{eke} had shown that the observed data of those time were
well approximated by such dependence with $\delta_c=1.686$, as could be 
inferred from the dynamics of the collapse of spherically symmetrical
cloud for the flat model of the Universe without cosmological constant.
The conclusion drawn by these authors regarding the weak dependence of
the $\delta_c$ value on 3D-space curvature and cosmological constant
paved the way for the canonical value $\delta_c=1.686$ to be used at arbitrary values
of 3D-space curvature and cosmological constant without the analysis of errors considered.
However, the data of the apparent magnitude - redshift testing for  Supernovae Ia become 
available since 1998 (see references in [10-12]) and
suggest that the Universe is expanding with acceleration. Thus the most plausible model of
the Universe rather strongly deviates from the canonical model with a zero 3D curvature and
zero cosmological constant.
Most probably, the matter density (dark matter plus baryons) is only
20-30\% of the critical density value, and the rest 70-80\% is a dark energy
(cosmological constant, vacuum energy, or quintessence). This requires
more careful approach to applying of the Press-Schechter formalism for  
determination of concentration, mass function and X-ray temperature function
of rich clusters of galaxies in different cosmological models for
their comparison with observed data.

The paper is aimed to:

\noindent - analyze the dynamics of the collapse of spherical dust-like
 cloud in models with arbitrary values of 3D curvature and cosmological constant;  

\noindent - to study dependence of $\delta_c$, the threshold value of the initial density perturbation 
 amplitude, on the above mentioned parameters,   
 as well as dependence of the final density 
 fluctuation after establishment of dynamical equilibrium, $\Delta_c$;

\noindent - construct approximations of these dependences on the values of the
 3D-space curvature and cosmological constant; 

\noindent - study deviation of the mass function and X-ray temperature of rich clusters of 
 galaxies employing real and canonical values of $\delta_c$ and $\Delta_c$
 (1.686 and 178 respectively)  for the models being analyzed in modern literature.

The analyses of the dynamics of the collapse of spherical
dust-like cloud for models with zero 3D-space curvature
and arbitrary values of cosmological constant as well as for generalized models,
with non-zero curvature, were carried out in   \cite{eke,lokas}.
  
In this paper we study the dynamics of the collapse of spherical
perturbations for models with arbitrary curvature and cosmological constant
parameters and analyze 
the influence of the dependence of critical amplitude on these parameters for the calculation of
mass function of rich clusters of galaxies.
Aiming at this, we will find solutions of appropriate equations, which describe the development 
of density perturbations at all evolutionary stages  
from initial linear one up to nonlinear collapse and virialization.

\section{The dynamics equation for a dust-like cloud when  $\Lambda\ne0$ and 
$k\ne0$}

In order to analyze the development of spherically symmetrical perturbations
in the Universe with a non-zero cosmological constant ($\Lambda\ne0$),
we consider the dynamics of dust like cloud in the comoving synchronous gauge.
Note that similar problem was first solved by Tolman in 1934
for the case  $\Lambda=0$ \cite{tolm}. 
We will consider the dynamics of spherical cloud before 
the appearance of first counterflows, that is trajectory intersections
of single particles (spherical shells).
At this stage, the dynamics of matter is well described by the continuum 
approximation, that is by the energy-momentum tensor of perfect fluid 
$$T_{ik}=(\varepsilon+P)u_iu_k-g_{ik}P,{   }(i,k=0,1,2,3),$$ 
where $\varepsilon=\rho c^2$  is energy density, $P$ is hydrodynamic pressure, 
$u_i$ -- 4D velocity components. 

The continuum approximation with $P=0$ allows to consider
perturbations in the gauge, which is simultaneously synchronous 
and comoving ($u_{\alpha}=0, \alpha=1,2,3$). In spatial 
spherical coordinates $R$, $\theta$ and $\varphi$ the metric of such Universe 
would take the following form ~\cite{land}

\nee{f1}{
ds^2=d\tau^2- e^{\lambda(R{,} \tau)}dR^2-r^2(R{,} 
\tau)(d\theta^2+\sin^2\theta d\varphi^2),}
where $\tau=ct$ is the time in units of length.
Einstein's equations with cosmological constant are so
$$R_{ik}-\frac{1}{2}g_{ik}R=\frac{8\pi G}{c^2}T_{ik}+g_{ik}\Lambda.$$
Calculating the tensor components $R_{ik}$ for the  metric (\ref{f1}) and
substituting them into Einstein's equations, and assuming $T_{\alpha \beta}=0$ for
our case (where $\alpha$ and $\beta=1,2,3$), we obtain the following set of equations:
\nee{f2}{
-\frac{ e^{-\lambda}}{r^2}(2r''r+r'^2-rr'\lambda')+ \frac{1}{r^2}(r\dot 
r\dot \lambda+{\dot r}^2+1)= 8\pi G\rho+\Lambda,}
\nee{f3}{
2\ddot rr+{\dot r}^2+1-r'^2 e^{-\lambda}=\Lambda r^2,}
\nee{f30}{
\ddot \lambda+2\frac{\ddot r}{r}+\frac{{\dot \lambda}^2}{2}+\frac{\dot \lambda \dot 
r}{r}-\frac{ e^{-\lambda}}{r}(2r''-\lambda'r')=2\Lambda,}
\nee{f4}{
\dot \lambda r'-2\dot r'=0.}
All other components are identically zeros.
By integrating the latest equation, the following relation is obtained:
$$ e^{\lambda(R,\tau)}=\frac{r'^2(R,\tau)}{1-f(R)},$$
where the value  $f(R)$ satisfies the condition $1-f(R)>0$.
Substituting it into (\ref{f3}) results in 
\nee{f10}{
2\ddot rr+{\dot r}^2-\Lambda r^2+f=0.}
Integrating of this equation gives
\nee{f5}{
{\dot r}^2-\frac{F(R)}{r}-\frac{\Lambda}{3}r^2=-f,}
where $F(R)$ -- time independent value.
This equation can be interpreted as an energy conservation equation 
for unit mass particle moving in the potential field 
$U=-({\Lambda}/{6})r^2-{F}/{(2r)}$ with total energy equal to 
$-{f}/{2}$. Multiplying (\ref{f2}) by $r'r^2$ and
using (\ref{f4}) and (\ref{f5}), we obtain
$$8\pi G\rho=\frac{F'}{r'r^2}.$$
Carrying out the integration of the latest equation it becomes
\nee{f6}{
F(R)=8\pi G\int\limits_0^R \rho r^2r'dR=2Gm,}
where $m$ -- total mass of the sphere with comoving radius $R$.
Since integrals are taken over the hypersurface of constant time, 
so $dr=r'dR$. 
By substitution 
$\rho=\overline{\rho}(1+\delta(R,\tau))$ (where 
$\delta(R,\tau)\equiv(\rho(R,\tau)-\overline{\rho}(\tau))/
\overline{\rho}(\tau)$ and $\overline{\rho}(\tau)$ - 
mean matter density in the Universe) into equation (\ref{f6}) we obtain
\nee{f11}{
F(R)= \frac{8\pi G}{3}\overline{\rho} 
r^3(1+\delta_r)=H_0^2\Omega_m\left(\frac{r}{a}\right)^3(1+\delta_r),}
where $$\delta_r(R,\tau)=\frac{3}{r^3}\int\limits_0^r \delta(y,\tau) y^2dy.$$
Also we have used here the definition for the mean matter density in
the Universe
$$\overline{\rho}(\tau)=\frac{3H_0^2\Omega_m}{8\pi 
G}a^{-3}(\tau),$$
where  $H_0$ is the current value of the Hubble constant, 
$\Omega_m$ -- the density of dust-like matter in units of the critical density of the Universe,
$a(\tau)$ -- scale factor of cosmological model with  metric
$$
ds^2=d\tau^2-a^2(\tau)\left[dR^2+\chi^2(R)
(d\theta^2+sin^2\theta d\varphi^2)\right],
$$
Тherefore
\nee{f7}{
\delta_r(R,\tau)=\frac{F(R)}{H_0^2\Omega_m^0}\left(\frac{a}{r}\right)^3-1,}
or 
\nee{f8}{
\delta_r(R,\tau)=(\delta_r(R,\tau_i)+1)\left(\frac{r_i}{a_i}\right)^3\left(\frac{a}{r}\right)^3-1,}
where the index $i$ denotes fixed values at a given period of time $\tau_i$.
In the absence of perturbations, the components of metric take the form  
$$
\begin{array}{l}
e^{\lambda(R,\tau)}\equiv a^2(\tau),\\
r(R,\tau)\equiv a(\tau)\chi(R).
\end{array} $$
In this case, the set of equations (\ref{f2}-\ref{f4}) 
will be reduced to the following 
\nee{f9}{
{\dot a}^2-\frac{8\pi G}{3}\overline{\rho}a^2-\frac{\Lambda}{3}a^2=-k,}
$$2\ddot a a+{\dot a}^2-\Lambda a^2+k=0,$$
$$\chi''(R)\chi(R)-{\chi'}^2(R)+1=0,$$
where $k=(1-{\chi'}^2(R))/\chi^2(R)$ -- curvature parameter.
The real solutions of the last equation are
$$\chi(R)=
\left(
\begin{array}{c}
\sin (\omega R)/\omega\\
R\\
\sh (\omega R)/\omega
\end{array}
\right),$$
where $\omega=\pi/R_0$, and $R_0$ -- radius of the Universe.
With them the curvature parameter takes values
$$k=\left(
\begin{array}{r}
1\\
0\\
-1
\end{array}
\right).$$
For closed models ($k=1$) $R_0$ is real number, in the case of open models ($k=-1$)
radius of the Universe is imaginary number.
The set of equations (\ref{f5}), (\ref{f8}) and (\ref{f9}) allows us to describe
the evolution of perturbations from early linear stage up to late nonlinear one. 
The initial conditions are following: 
$\delta_{r{,}i}=\delta_r(a_i)$ and $r_i=r(a_i)$.

\section{ Linearization of Tolman's model equations and their solutions}

At first, let us consider this set of equations for the case of small perturbations,
namely for $\delta\ll 1$. Then the metric components (\ref{f1})
may be written in the following way
\nee{f12}{
e^{\lambda(R,\tau)}=a^2(\tau)(1+h_{11}(R,\tau)),{   }
r(R,\tau)=a(\tau)\chi(R)(1+h_{22}(R,\tau)).}
Employing (\ref{f5}) and (\ref{f11}), one may represent the equation (\ref{f10}) 
in the form:
$$2\ddot rr-\frac{2\Lambda}{3}r^2+\frac{8\pi 
G}{3}\overline{\rho}(1+\delta_r)r^2=0.$$
For small perturbations ($h_{11}\ll1$, $h_{22}\ll1$) the last equation in 
the linear approximation acquires the form
\nee{f13}{
a^2\ddot h_{22}+2\dot a ah_{22}+\frac{4\pi 
G}{3}\overline{\rho}\delta_ra^2=0.}
In order to find the relation between the $h_{22}$ аnd $\delta_r$, let's
insert (\ref{f12}) into (\ref{f7}) and confine ourselves by the 1-st order values 
$$
\delta_r=\frac{F(R)}{H_0^2\Omega_m\chi^3(R)} 
\left(1-\frac{3h_{22}}{2}\right)-1.
$$
Switching onto the case of non-perturbed metric ($h_{22}\equiv0$), one obtains 
$$\delta_r={F(R)}/{(H_0^2\Omega_m\chi^3(R))} -1,$$ 
with $\dot\delta_r=0$. The presence of density perturbations in the matter 
distribution inevitably leads to respective perturbations in the metric, 
that is $h_{22}\ne0$. Therefore, the amplitude $\delta_r$ for $h_{22}\equiv0$ 
must to be identically 0. Hence
\nee{f14}{
F(R)=H_0^2\Omega_m\chi^3(R).}
In other words, the analysis of perturbation behavior is carried out by 
comparing space and velocity distributions of the same particles in the regions 
that contain equal mass $m$ within the same comoving $R$. 
The equation for the growth of matter 
perturbations  (\ref{f7}) when (\ref{f14}) is assumed, becomes the next
$$\delta(R,\tau)=\left(\frac{a\chi(R)}{r}\right)^3-1.$$
This equation is often used for the correction of linear power spectrum 
for the nonlinear evolution of the perturbation ~\cite{core}.
Taking it into account one may obtain
\nee{f19}{
h_{22}=-\frac{1}{3}\delta_r(R{,}\tau).}
In fact, this condition let to single out from the arbitrary
initial conditions those ones which determine a scalar perturbation mode -- 
of the most interest in cosmological studies.
The substitution of the latest equation into (\ref{f13}) provides the following
linear equation for the evolution of the small perturbation (see also ~\cite{pibls})
\nee{pib}{
\ddot\delta_r +2\frac{\dot a}{a}\dot\delta_r-4\pi 
G\overline{\rho}\delta_r=0.}
The general solutions can be expressed by the relative difference in background 
densities of the two similar models which differ in integration constants
$$\delta_{r{,}j}\sim\frac{1}{a}\frac{\partial a}{\partial \alpha_j},$$
where $\alpha_j$ -- integration constant of the cosmological equations 
for background.

To find the solutions of (\ref{f9}) we need the initial value of the 
scale factor $a_i\equiv a(\tau_i)$ for some arbitrary time $\tau_i$. 
In particular, for the start time $\tau=0$ we fix
the value of scale factor $a(\tau=0)=0$. Due to this we can obtain the value of scale 
factor from (\ref{f9}) for any time and in particular for this moment $\tau_0$.
The current time $\tau_0$ is obtained from Hubble constant. 
This constant is denoted as $H_0$ at current time, and the scale factor as $a_0$. 
In the case of non-zero curvature
$$
a_0=\sqrt{\frac{k}{\frac{8\pi G}{3} \overline{\rho}_0+\frac{\Lambda}{3} - H_0^2
}}.
$$
Values of $\overline {\rho}_0$ and $\Lambda$ at present time are defined through physical 
coordinates and not dependent on value of $a_0$.
In the case of zero curvature the scale factor is indefinite and could by taken 
deliberately. 
Apparently, the physical sense  
is contained in the ratio $a(\tau_1)/a(\tau_2)$, where $\tau_1$ and $\tau_2$ 
any moments of time you choose, not in a base value of scale factor.
Let us reduce the scale factor to its present value $a_0$. 
So, when noting this rate as $\tilde a=a/a_0$, the equation (\ref{f9}) takes the form 
$${\dot {\tilde a}}^2-\frac{8\pi G}{3}\overline{\rho}{\tilde a}^2-
\frac{\Lambda}{3}{\tilde a}^2=-\tilde k,$$
where $\tilde k=k/a_0^2$ takes continuous values. Below we denote $\tilde k$ by $k$,
considering it as continuous quantity, and $\tilde a$ by $a$, putting scale factor 
at the present time equal to unity. The quantities $\tau_0$ and $a_0$, or, what is the 
same, $\tau_0$ and $k$, are integrals of cosmological equation. 
Thus, we obtain two solutions of equation (\ref{pib})
$$\delta_{r{,}1}\sim\frac{1}{a}\frac{\partial a}{\partial \tau_0},{    } 
\delta_{r{,}2}\sim\frac{1}{a}\frac{\partial 
a}{\partial k}.$$
According to the above definition
\nee{w1}{
\delta_{r{,}1}\sim\frac{\overline{\rho}(\tau{,}k{,}\tau_0+d\tau_0) 
-\overline{\rho}(\tau{,}k{,}\tau_0)}{ 
\overline{\rho}(\tau{,}k{,}\tau_0)}, 
\delta_{r{,}2}\sim\frac{\overline{\rho}(\tau{,}k+dk{,}\tau_0) 
-\overline{\rho}(\tau{,}k{,}\tau_0)}{ \overline{\rho}(\tau{,}k{,}\tau_0)}.}
Using the integral of the equation (\ref{f9}), the last expressions can be 
represented as
$$\delta_{r{,}1}\sim\frac{1}{a} X^{1/2}, {   }
\delta_{r{,}2}\sim\frac{1}{a}X^{1/2}\int\limits_0^a X^{-3/2}da,$$
wherе $X=\frac{\Lambda}{3}a^2+\frac{8\pi G}{3}\overline{\rho} 
a^2-k.$
We are interested in the growing solution $\delta_{r{,}2}$ so far. 
Let's introduce the designation ~\cite{pibls}
\nee{f15}{
D(a)= \frac{5H_0^2\Omega_m}{2a}X^{1/2}\int\limits_0^a X^{-3/2}da}
Then the exact linear solution for the increasing mode of matter density 
perturbation is
\nee{f16}{
\delta_r^l(a)= \frac{\delta_r(a_i)}{D(a_i)}D(a).}
In view of the fact that perturbation at the initial stages was very small, 
one may assume  
\nee{f17}{
\delta_r(a) \mathop{\simeq}_{a\ll1} \delta_r^l(a).}
When $a\ll1$, one may expand (\ref{f16}) into series over a small parameter $a$, so that
\nee{f20}{
\delta_r(a)\simeq\delta_r^l(a) \simeq A(a+O(a^2)),}
where
\nee{f21}
{A=\frac{\delta_r(a_i)}{D(a_i)}.}

\section{ Parameters $\delta_c$ and  $\Delta_c$, and their dependences on $\Lambda$ and 
$k$ }

One of the important values used in cosmology for analyzing conditions of the  
 large-scale structure of the Universe formation is the initial amplitude of matter
perturbations which collapse at the given moment of time.
It is called threshold amplitude and denoted by $\delta_c$ and depends both on the parameters of cosmological 
model and on the time moment of collapse 
$\tau_{col}$ ($\delta_c\equiv\delta_c(\Omega_m,\Omega_{\Lambda},\tau_{col})$). 
In the framework of linear theory of perturbations the increasing solution is
described by the expression (\ref{f16}). To find the solution that would describe the case for
$\delta\ge1$, let's use the fact that the increasing mode of scalar perturbation is determined
by the difference between the models with different curvature and the same start time
$\delta_r=\delta_{r{,}2}$ (\ref{w1}). This makes possible to extract the increasing part
out of the general solution (\ref{f7}) by use of the time synchronization of integrals 
of equations (\ref{f5}) and (\ref{f9}) respectively
$$\int\limits_0^a\frac{da}{\sqrt{\frac{\Lambda}{3}a^2+\frac{8\pi G}{3}\overline{\rho} 
a^2-k }} 
=\tau-\tau_0,{  }
\int\limits_0^{r(a)}\frac{dr}{\sqrt{\frac{\Lambda}{3}r^2+\frac{F}{r}-f 
}}=\tau-\tau_0.$$
Equating them gives
\nee{f18}{
\int\limits_0^a\frac{da}{\sqrt{\frac{\Lambda}{3}a^2+\frac{8\pi G}{3}\overline{\rho} 
a^2-k }} 
=\int\limits_0^{r(a)}\frac{dr}{\sqrt{\frac{\Lambda}{3}r^2+\frac{F}{r}-f 
}}.}
At the moment when perturbation turns around, i.e. when $\dot r=0$, $r$ 
reaches its maximal value $r=r_{ta}$, where $r_{ta}=r(a_{ta})$ and can be obtained
from the equation
$$\frac{\Lambda}{3}r_{ta}^3-fr_{ta}+F=0.$$
At the same moment the right-hand  integrand in the equality (\ref{f18}) takes
an infinitely large value, but the integral remains
an finite quantity. So, let's rewrite the  (\ref{f18}) to the form
$$\int\limits_0^{a_{ta}}\frac{da}{\sqrt{\frac{\Lambda}{3}a^2+\frac{8\pi 
G}{3}\overline{\rho} a^2-k }} 
=PV\int\limits_0^{r_{ta}}\frac{dr}{\sqrt{\frac{\Lambda}{3}r^2+\frac{F}{r}-f 
}},$$
wherе $PV$ means principal integral value.

As we already mentioned, the equation (\ref{f5}) is equivalent to the energy conservation
equation for the massive particle motion in given spherically symmetrical potential field. 
It is well known, that in the case of negative total energy (that is when $f>0$) such motion
is symmetrical in time with respect to $\dot r=0$ moment, thus 
$$\tau_{col}=2\tau_{ta},$$
wherе $\tau_{col}$ -- moment of the collapse of the cloud. We have assumed here that
the expansion of cloud started at zero moment of time. This equality can be 
rewritten in the following way:
$$\int\limits_0^{a_{col}}\frac{da}{\sqrt{\frac{\Lambda}{3}a^2+\frac{8\pi 
G}{3}\overline{\rho} a^2-k }} 
=2PV\int\limits_0^{r_{ta}}\frac{dr}{\sqrt{\frac{\Lambda}{3}r^2+\frac{F}{r}-f 
}},$$
where $a_{col}=a(\tau_{col})$.
Taking into account (\ref{f14}), one obtains
\nee{f23}{
\int\limits_0^{a_{col}}\frac{\sqrt{a}da}{\sqrt{\Omega_{\Lambda}a^3+\Omega_ka+\Omega_m 
}} 
=2\int\limits_0^{x_{ta}}\frac{\sqrt{x}dx}{\sqrt{\Omega_{\Lambda}x^3+\Omega_fx+\Omega_m
}},}
wherе $\Omega_{\Lambda}={\Lambda}/(3H_0^2)$, $\Omega_m={8\pi 
G{\overline{\rho}}_0}/{(3H_0^2)}$, $\Omega_k={-k}/{H_0^2}$, 
$\Omega_f=-f/(H_0^2\chi^2(R))$, $x=r/\chi(R)$, $x_{ta}=r_{ta}/\chi(R)$.

Setting the time of collapse and using the last equality, one can find the value
$\Omega_f$ for the perturbation collapsing at the moment $a_{col}$.
Substituting (\ref{f12}) with (\ref{f19}) and (\ref{f14}) into 
(\ref{f5}) and confining ourselves to the 1-st order quantities:
$$
{\dot a}^2a\frac{d\delta_r}{da}+\left({\dot 
a}^2+\frac{4\pi G}{3}\overline{\rho} a^2-\frac{\Lambda}{3}a^2\right) 
\delta_r=\frac{3}{2}\left(\frac{f}{\chi^2}-k\right).
$$
Considering the equality only at small $a$, substituting (\ref{f20}) into this 
equation and writing out components at zero power of $a$ gives:
$$\frac{20}{3}\pi 
G\overline{\rho}_0A=\frac{3}{2}\left(\frac{f}{\chi^2}-k\right),$$
Hence
$$A=\frac{3(\Omega_k-\Omega_f)}{5\Omega_m}.$$
Taking into account (\ref{f15}), (\ref{f16}) and (\ref{f21}) we obtain

\nee{f24}{
\delta_r^l(a) = \frac{3(\Omega_k-\Omega_f)}{5\Omega_m}D(a).}

Thus, the initial amplitude of increasing mode of the perturbation is
determined by the difference of curvature parameters of the background 3D-curvature parameter $\Omega_k$
and local perturbation $\Omega_f$ one. Fixing the time of collapse $\tau_{col}$ in 
arbitrary way and determining the value $\Omega_f$ from (\ref{f23}), we substitute 
it in (\ref{f24}) and obtain the sought value of $\delta_c$
\nee{form1}{
\delta_c(a_{col},a)=\frac{3(\Omega_k-\Omega_f(a_{col}))}{5\Omega_m}D(a),}
where $a_{col}$ -- scale factor at the moment of collapse, 
$a$ -- scale factor at the moment for which
recalculation of the critical amplitude
is done according to the linear theory of perturbations.
In the case of $\Omega_k=0$, after appropriate redesignations,
the above-stated formula is equivalent to the similar value obtained by ~\cite{eke}. 
For the simplest case of $\Omega_m=1$ and $\Omega_k=0$ it coincides 
with the well known one from literature (see for example ~\cite{om1,sim}), 
since, as it may be easy shown, the critical amplitude will take on the form  
$$\delta_c(a_{col},a)=\frac{3}{5}\left(\frac{3\pi}{2}\right)^{2/3}
\frac{a}{a_{col}}\simeq 1.68647\frac{a}{a_{col}},$$
and at $a=a_{col}$ it has the canonical value.
 
\begin{figure}[hbt]
\begin{center}
\epsfxsize=20cm
\epsffile{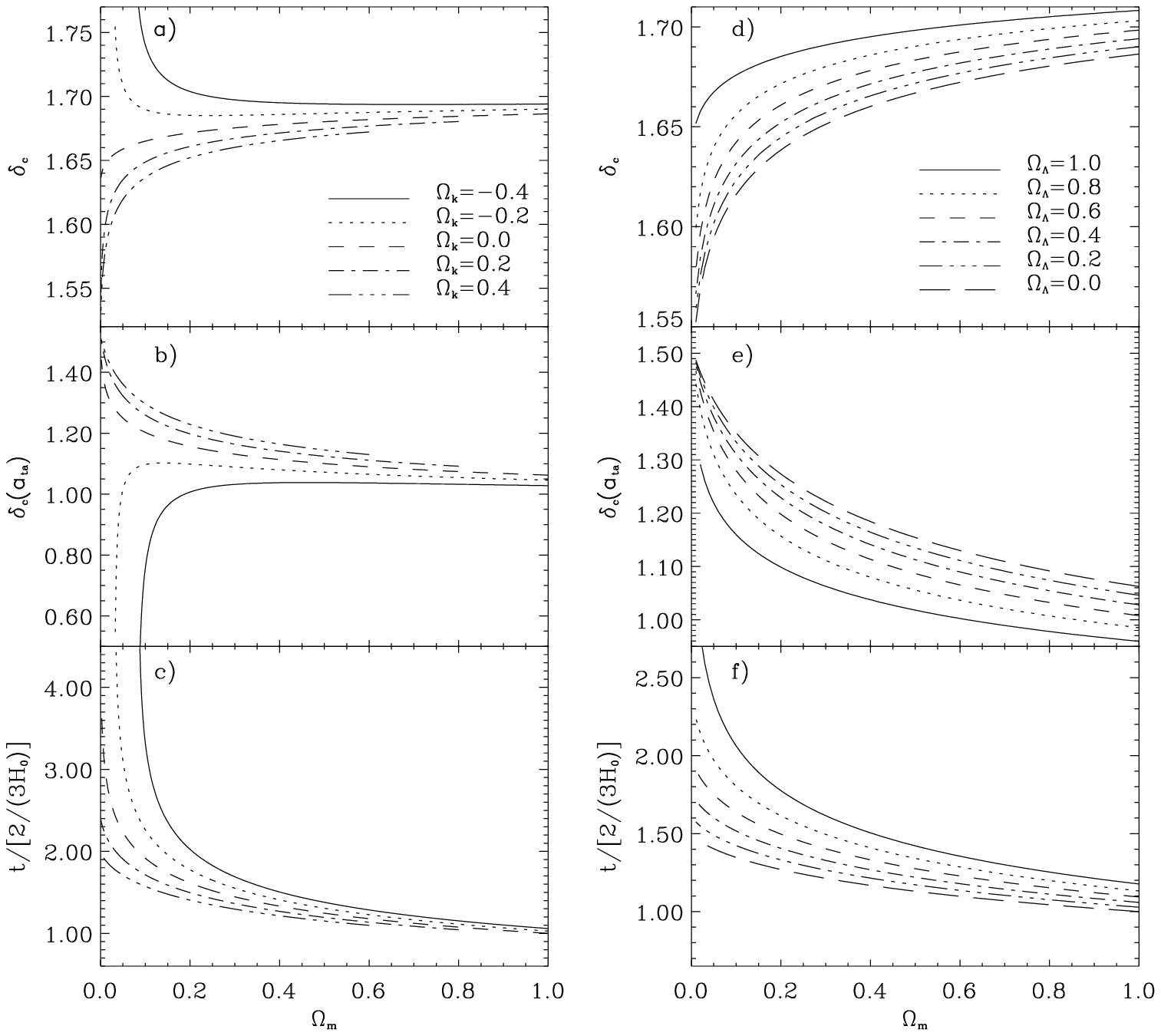}
\end{center}
\caption{\small Dependence of the threshold value of the amplitude of
density fluctuation collapsed at the current epoch ($\delta_c$) 
and calculated for the moment of collapse (a, d), 
and for the moment of turn around of expansion of the spherical cloud (b, e) on 
the matter abundance $\Omega_m$ for different $\Omega_k$ and $\Omega_{\Lambda}$.  
Also, dependence of the age of the Universe for 
corresponding cosmological models expressed in units of the age of the Universe 
for the canonical model with $\Omega_m=1$ and $\Omega_k=\Omega_{\Lambda}=0$ are 
presented below (c, f).
}\label{multy_d}
\end{figure}

Fig.~\ref{multy_d} shows dependence of $\delta_c$ on $\Omega_m$
for the models with fixed values of curvature parameters $\Omega_k$ (a) and 
cosmological constant $\Omega_{\Lambda}$ (d).
In the models with positive values of cosmological constant the condition
of positive curvature of local perturbation is not sufficient 
for its ever collapse since the positive cosmological constant counteracts to
deceleration at the stage of expansion of perturbation.
The local curvature ($f/\chi^2(R)$) needed for the collapse
must exceed some critical value determined from the condition that there exists 
the moment of turn around of perturbation region where $\dot r=0$. 

The  magnitude $\dot r^2$  according to (\ref{f5}) acquires the 
minimum at $r=r_{extr}\equiv(\Omega_m/(2\Omega_{\Lambda}))^{1/3}$. 
Then
$$\dot r_{min}^2=\dot r^2(r_{extr})=3(\Omega_m/2)^{2/3}\Omega_{\Lambda}^{1/3}+
\Omega_f.$$ 
In order the turn around moment to exist there must be $\dot r_{min}^2\le 0$ (only at this condition
there is a moment when $\dot r =0$). That means
$$\Omega_f \le - 3(\Omega_m/2)^{2/3}\Omega_{\Lambda}^{1/3}.$$

\begin{figure}[hbt]
\begin{center}
\epsfxsize=14cm
\epsffile{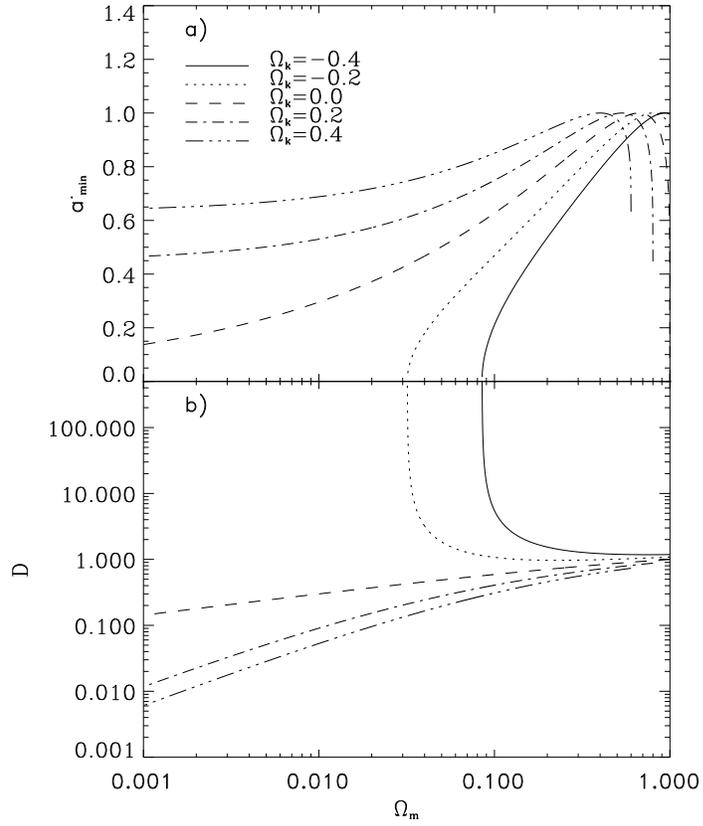}
\caption{\small 
Dependence of the smallest expansion rate of the Universe (a) 
and an growth factor $D(a=1;\Omega_m)$ (b) on cosmological parameters.}
\label{multy_Da}
\end{center}
\end{figure}

As it was already shown above, the initial value of  perturbation
is determined by the difference  between the  background curvature and the local
curvature of the perturbation region (see (\ref{f24})). For the models with negative 
curvature ($\Omega_k>0$), the initial value of the density perturbation amplitude
of collapsing cloud
always must exceed some critical value, because the collapse region must have a 
positive value of curvature.
Moreover, the larger $\Omega_k$ is the larger value of the amplitude of perturbation takes
at the turn around of its expansion, as calculated within the framework of 
linear theory (see (\ref{form1}) and  Fig.\ref{multy_d}b and  e). 
Among background models with a positive curvature ($\Omega_k<0$), 
one can find models with considerable slow-down of their expansion. 
Moreover, the perturbation amplitude needed for collapse can be infinitesimal 
(Fig.\ref{multy_d}b). 
However, after recalculation according to the linear theory for the present time 
(Fig.\ref{multy_d}a), the initially small perturbation amplitude tends to large values.  
This is due to the fact that these models have long period of time when the rate 
of expansion of the background model was close to zero $\dot a\simeq0$ 
(Fig.~\ref{multy_Da}a). 
Both for the perturbation region and background, the minimum value for the rate of 
expansion, according to (\ref{f9}),  is reached when 
$a=a_{extr}=(\Omega_m/(2\Omega_{\Lambda}))^{1/3}$.
Similarly,
$$\dot a_{min}^2=\dot a^2(a_{extr})=3(\Omega_m/2)^{2/3}\Omega_{\Lambda}^{1/3}+
\Omega_k.$$ 

In contrast to the perturbation region, the smallest rate of expansion for background 
models with $\Omega_m<2\Omega_{\Lambda}$ ($a_{extr}<1$) have to be greater than zero. 
Otherwise, we would observe now a stationary or collapsing model of the Universe. At 
the moment of the minimum rate of expansion, as it could be easily shown, the 
acceleration equals zero. Nearly zero rate of expansion and its invariability 
with time leads to the appearance of a long-lasting stage in the evolution of the 
Universe, where the value for the scale factor is almost constant.
According to (\ref{pib}), perturbations at this stage grow exponentially 
$$\delta\sim e^{\alpha \Delta \tau},$$
where $\alpha=\sqrt{4\pi G\overline{\rho}_0}a_{extr}^{-3/2}$, а $\Delta\tau$ -- 
duration of a considerable slow-down in the background expansion 
(in units of the Hubble constant). 
The closer the minimum rate of expansion is to zero (see Fig.~\ref{multy_Da}a),
the longer stage is and the increase of amplitude is greater  
(see Fig.~\ref{multy_Da}b).
The characteristic time for this stage can be obtained on the basis of (\ref{f9})
$$\Delta\tau\simeq \frac{a_{extr}}{\dot a_{min}}\epsilon,$$
where $\epsilon=\Delta a/a$ -- an allowable relative change in the scale factor. 
At the infinitesimal value of $\epsilon$ and nearly-zero value of $\dot a_{min}$
the time of exponential increase of perturbations $\Delta \tau$ may take arbitrary large values.
Correspondingly, an increase of the perturbation amplitude for this period of time is large as well. 
Thus, cosmological models with a substantial slow-down in expansion are gravitationally unstable. 

\begin{figure}[hbt]
\begin{center}
\epsfxsize=20cm
\epsffile{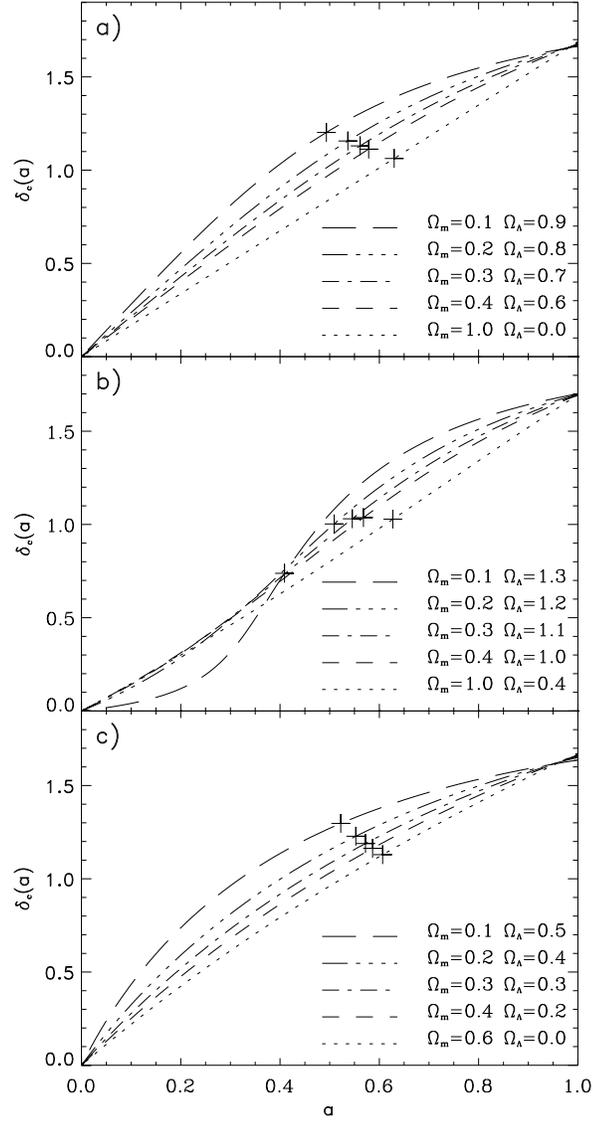}
\end{center}

\caption{\small Dependence  of the 
$\delta_c$ on the scale factor for 
set of models with zero curvature (a), for models 
with $\Omega_k=-0.4$ (b) and $\Omega_k=0.4$ (c).
Crosses denote moments of turn around of expansion of the perturbation region. 
}\label{multy_dc}
\end{figure}

In models with $\Omega_k\ge0$ such slow-down is essentially less, resulting in an inverse relation between $\delta_c(a_{col})$
and $\delta_c(a_{ta})$:  $\delta_c(a_{col})/\delta_c(a_{ta})$ decreases with decrease of 
$\Omega_m$.
Such behavior is caused by decrease of the rate of perturbation 
growth at the final stage of the collapse of perturbations  due to an increased 
influence of the cosmological constant and curvature of the Universe 
(Fig.~\ref{multy_dc}). 

\begin{figure}[hbt]
\begin{center}
\epsfxsize=14cm
\epsffile{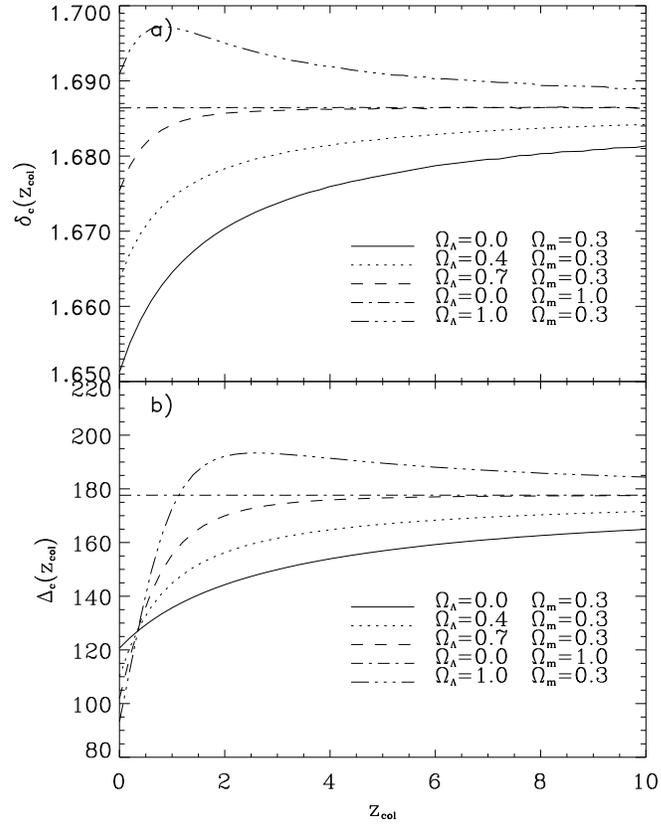}
\end{center}
\caption{\small Dependence of threshold linear density fluctuation $\delta_c$  (a) 
and overdensity after virialization $\Delta_c$ (b) on the moment of collapse $z_{col}$
for different values of the matter density  and cosmological constant.
}\label{multy_dz}
\epsfxsize=9cm
\end{figure}

Dependence of the critical amplitude  $\delta_c\equiv\delta_c(z_{col})$ recalculated 
to the moment of collapse, allows to evaluate a change of influence of the 
cosmological parameters on the dynamics of the growth of perturbations in the
expansion Universe (Fig.~\ref{multy_dz}a). 
It is clear that the influence of the cosmological constant on $\delta_c$ with 
increasing of $z$ diminishes as it is also substantiated by graphs. 

\begin{figure}[hbt]
\begin{center}
\leavevmode
\epsfxsize=14cm
\epsffile{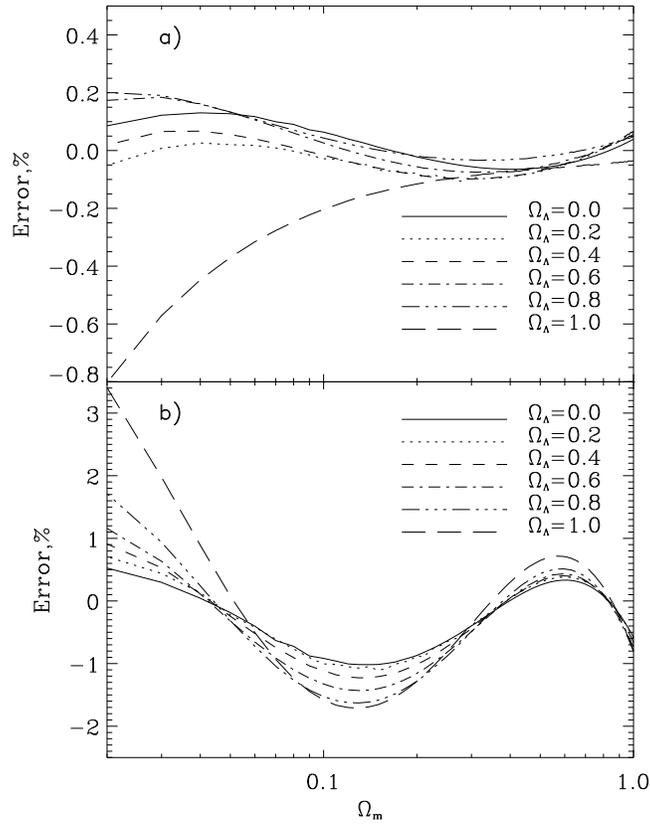}
\end{center}
\caption{\small Dependence of the relative error of the analytical approximation 
for the values of $\delta_c$ (a) and $\Delta_c$ (b) on the parameters of the 
cosmological model.
}\label{multy_er}
\end{figure}

Dependence of $\delta_c(a_{col}=1,a=1)$ on $\Omega_m$ and $\Omega_{\Lambda}$ can be 
expressed by the following approximate formula: 
$$\delta_c=A-B\ln(\Omega_m+C),$$
where 
$$\begin{array}{l}
A=1.687+0.021\Omega_{\Lambda},\\
B=0.01568(\Omega_{\Lambda})^{1.774}-0.03061,\\
C=0.00195-0.01135\Omega_{\Lambda}+0.00382(\Omega_{\Lambda})^2.
\end{array}
$$
The numerical values of the coefficients are found by the method of best-fit adjustment.
The accuracy of approximation for different parameters of the model are shown 
in Fig.~\ref{multy_er}a.

Note that the real amplitude  of perturbation $\delta$, calculated according to (\ref{f7}), 
is infinitely large at the moment of collapse. In fact, at the final stage of 
collapse of a cloud the energy is redistributed with subsequent equilibrium due to the 
short-range particle interaction and a finite target parameter. 
At very small scales, the gravitational force is counteracted by the short-range 
interaction that is determined by the nature of the particles in a cloud. 
At large scales, the key role in formation of stationary structures plays 
the balance of inertial and gravitational forces. 
According to the virial theorem,  a system of particles, approaching its dynamical 
equilibrium, has the following kinetic energy per unit mass  ~\cite{lahav}
$$T=\frac{1}{2}\sum_i v_i^2 = \frac{1}{2}\sum_i {{\overrightarrow{r}}_i} 
\overrightarrow{\nabla}_{{\overrightarrow{r}}_i} U.$$
In our case: $U=U_{\Lambda}+U_{\rho}=-\frac{\Lambda}{6}r^2-\frac{F}{2r}$,
so
$$T=-\frac{\Lambda}{6}r^2+\frac{1}{2}\frac{F}{2r}=U_\Lambda-\frac12U_\rho.$$
The moment when this equality is true we shall call the moment of virialization 
and denote with an index '$vir$'. 
Let's make use of the conservation law expressed in the form:
$$T+U_{\Lambda}+U_{\rho}=U_{\Lambda,ta}+U_{\rho,ta}.$$
At the moment of virialization, the last equality takes on the form of ~\cite{lahav}

$$2U_{\Lambda,vir}+\frac{1}{2}U_{\rho,vir}=U_{\Lambda,ta}+U_{\rho,ta},$$
or another form

$$4\Omega_{\Lambda}x_{vir}^3+2\Omega_f x_{vir}+\Omega_m=0,$$
where $x_{vir}=r_{vir}/\chi(R)$.

It would be reasonable to introduce here a quantity to characterize 
the relative divergence of density taking into account virialization process. 
This quantity, called as overdensity after virialization, is defined in the following way:
$$\Delta_c=\frac{\rho_{vir}}{\rho_c}\left(a_{col}\right)=
\frac{\Omega_mH_0^2}{x_{vir}^3H^2(a_{col})}.$$
In the case of the canonical model ($\Omega_{\Lambda}=0$, $\Omega_m=1$, $\Omega_k=0$)
$$\Delta_c=18\pi^2\simeq 178.$$

\begin{figure}[hbt]
\begin{center}
\epsfxsize=14cm
\epsffile{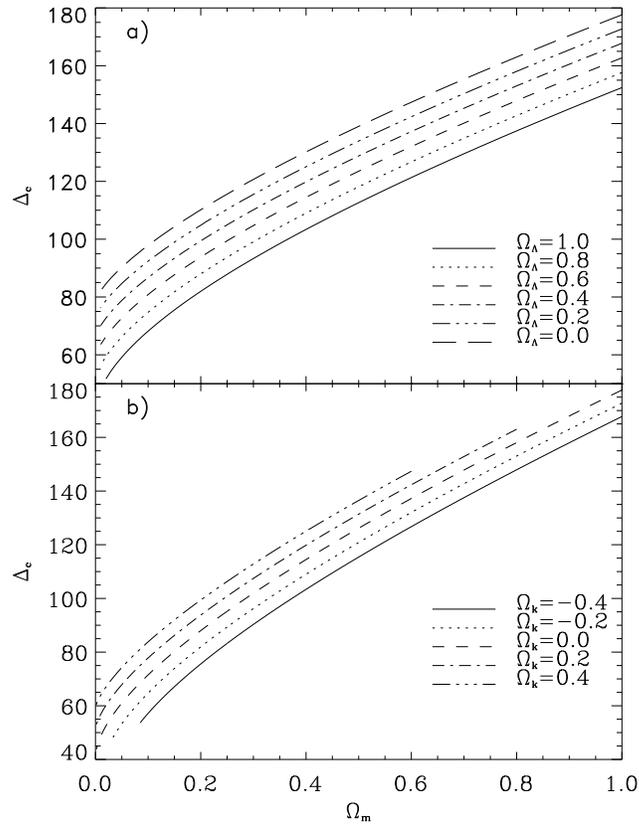}
\end{center}
\caption{\small Dependence of the overdensity after virialization $\Delta_c$
on the matter abundance $\Omega_m$ for different fixed values of a 
cosmological constant (a) and curvature (b).}
\label{multy_Ddc}
\end{figure}

Dependence of $\Delta_c$ on cosmological parameters is shown in Fig.~\ref{multy_Ddc} and 
Fig.~\ref{multy_Ddc}b. 
At scales of stationary clusters of galaxies, the density of the energy that is 
connected with the cosmological constant is far less than the energy of the matter, 
hence it has a subtle influence on the final value of the radius  $r_{vir}$.  
In this case, we may assume with a high certainty that 
$r_{vir}\simeq r_{ta}/2$. 
Thus, the final amplitude of the virialized perturbation is completely determined  
by scale of turn around $x_{ta}\equiv r_{ta}/\chi(R)$. 
When the relative density of the matter $\Omega_m$ increases, the perturbation 
amplitude grows nearly proportionally to it.
Although for the models with small values of $\Omega_m$ this linearity is broken because of
the increase in the role of cosmological constant (see Fig.~\ref{multy_Ddc}a 
and Fig.~\ref{multy_Ddc}b). 

The behavior of overdensity  $\Delta_c$ could
be explained by analyzing the influence of every parameter on its
dynamics. As far the matter component $\Omega_m$ makes the process
of expansion more slow, so the value of $r_{ta}$ should be decreasing
 when this component's content increase (one of the components, either 
 $\Omega_\Lambda$ or $\Omega_k$ we will assume to be fixed).
So that the value $\Delta_c(a_{col}=1) = \Omega_m/x_{vir}^3$ will be growing.
In the same way, since component $\Omega_\Lambda$ accelerates the expansion,
$r_{ta}$ will be growing. This leads to decrease in amplitude $\Delta_c$ when
$\Omega_m$ or $\Omega_k$ is held fixed.

As it was shown above, the relative amplitude of perturbation depends
linearly on the difference of curvatures, $\Omega_k-\Omega_f$. The local curvature
parameter, $\Omega_f$, depends on parameters of background model for the fixed 
collapse moment. However, because of slow variation in $\Omega_f$ against $\Omega_k$ at
the $\Omega_m$ or $\Omega_\Lambda$ being fixed, the change in value of
$\Omega_k-\Omega_f$ will be determined by the change in value of curvature 
parameter $\Omega_k$.
Therefore, if $\Omega_k$ increase, then $\Delta_c$ do the same. These relations of
$\Delta_c$ dependence to the cosmological parameters are illustrated in 
Figs.~\ref{multy_Ddc}a,b. 

As it was noted before, the influence of $\Omega_\Lambda$ and $\Omega_k$
on the dynamics of Universe at early stages was diminished by expansion. This makes 
the variation of $\Delta_c$ value within two arbitrary models smaller (see Fig.~\ref{multy_dz}b).
While we are moving to the earlier stages, the values $\delta_c$ and $\Delta_c$ 
would converge to the their canonical values 
 $3/5 (3\pi/2)^{2/3}\simeq 1.686$ and
$18\pi^2\simeq178$ correspondingly (Fig.~\ref{multy_dz}a,b).

The analytical approximation for the above mentioned value has the following form 
$$\Delta_c=A - B\ln(\Omega_m + C),$$
where
$$
\begin{array}{l}
A=110.89 - 12.91\Omega_{\Lambda} + 7.92(\Omega_{\Lambda})^2 - 0.03/\Omega_m,\\
B=-113.28 + 6.84\Omega_{\Lambda} + 7.13(\Omega_{\Lambda})^2 ,\\
C=0.787 - 0.115\Omega_{\Lambda} - 0.093(\Omega_{\Lambda})^2 .
\end{array}
$$
Numerical values of the coefficients are determined using the method of  best-fit 
adjustment. 
Accuracy of the approximation for different models is shown in Fig.~\ref{multy_er}b.

\section{ Mass function of rich clusters of galaxies}

\begin{figure}[hbt]
\begin{center}
\leavevmode
\epsfxsize=20cm
\epsffile{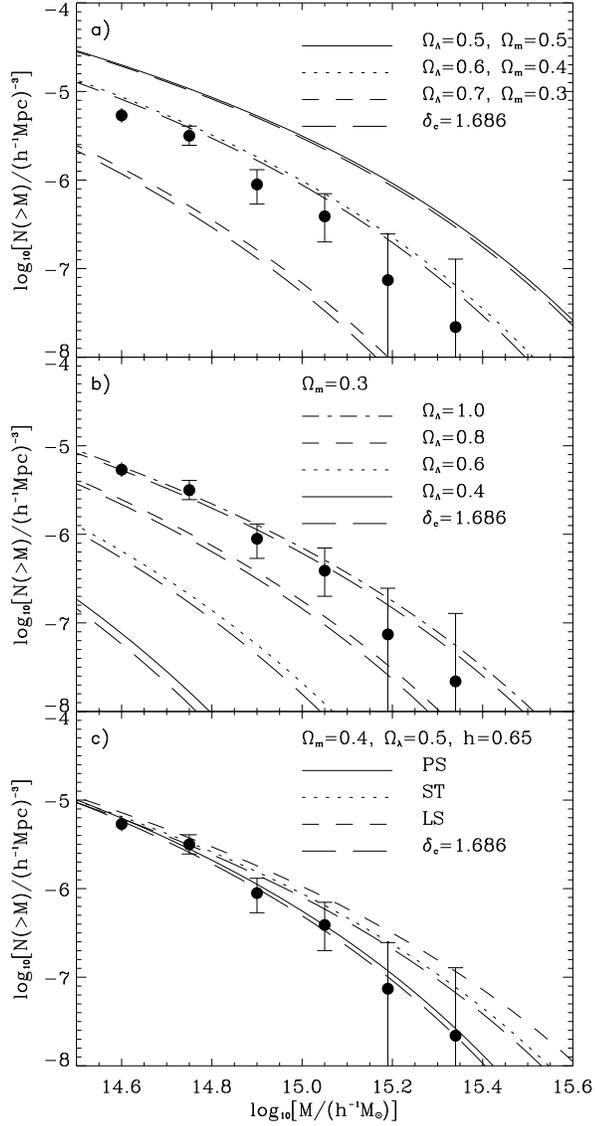}
\end{center}
\caption{\small Mass function of rich clusters of galaxies
for models with a zero curvature (a),
fixed value $\Omega_{\Lambda}$ (b)
and the better model with  $\Omega_{k}=-0.1$ (c) calculated according 
to the Press-Schechter (PS) \cite{ps}, Sheth-Tormen (ST) 
\cite{ST}  and Lee-Shandarin (LS) \cite{LS} $h=0.65$ approaches.} 
\label{multy_mf}
\end{figure}

\begin{figure}[hbt]
\begin{center}
\leavevmode
\epsfxsize=14cm
\epsffile{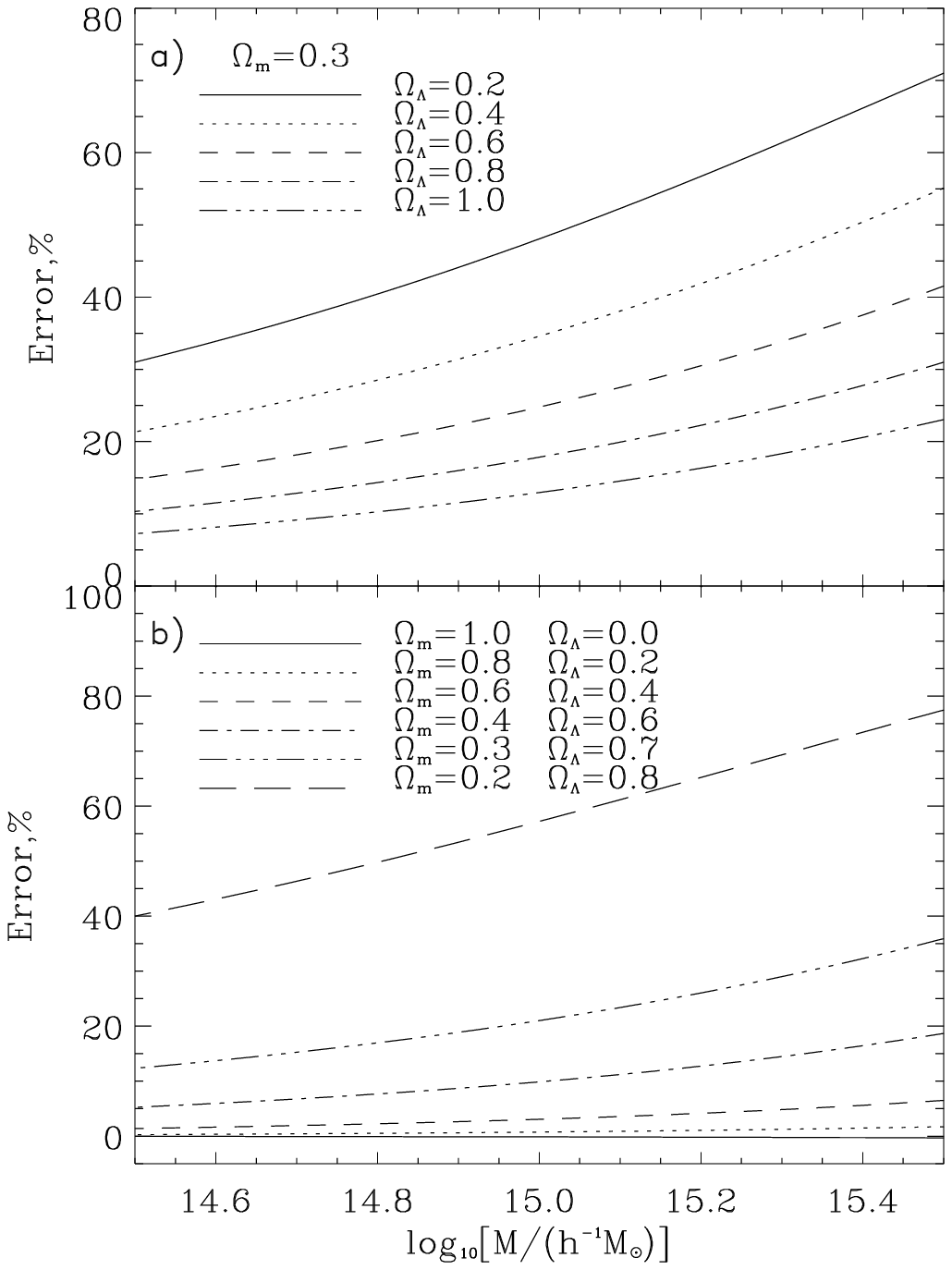}
\end{center}
\caption{\small 
Relative difference between concentrations of rich clusters of galaxies
calculated according to the Press-Schechter formalism. 
The difference arises from calculations with real and canonical values of $\delta_c$
for fixed values of $\Omega_{\Lambda}$ (a) and for models with a zero-curvature (b).
}\label{Multy_mfer}
\end{figure}

Rich clusters of galaxies are supposed to have been formed at the highest peaks of the 
matter density fluctuations -- which, as we know, are almost spherically symmetrical. 
That's why these formations can be described using the spherical collapse model. 
One of the most important features of the large-scale structure of the Universe at the 
scales $\sim10$ Мpc is a mass function that determines the number of gravitationally 
bound systems with their masses exceeding the given one. 
To calculate this distribution in the case of spherically symmetrical collapse it is 
convenient to use the Press-Schechter formalism.
The amplitude of initial perturbations ($\delta\ll1$)  
is assumed to be distributed according to the Gauss law 
$$P(R,\tau)=\frac{1}{\sqrt{2\pi}\sigma(R,\tau)}e^{-\frac{\delta^2}{2\sigma^2(R,\tau)}},$$
where
$$\sigma^2(R,\tau)=<\delta_s^2>=\frac{1}{2{\pi}^2}\int d^3k |{\delta_k}|^2 |{W(kR)}|^2.$$
A fraction of collapsed perturbations equals 
$$F(R,\tau)=\int_{\delta_c}^{\infty}\frac{d\delta}{\sqrt{2\pi}\sigma(R,\tau)}
e^{-\frac{\delta^2}{2\sigma(R,\tau)^2}}.$$
The number of the objects collapsed per unit volume with their masses within $dM$, 
according to the Press-Schechter approach constitutes  ~\cite{ps}
\nee{w2}{
n(M,\tau)dM=-2\frac{\overline{\rho}}{M}\frac{\partial F}{\partial
R}\frac{dR}{dM}dM=-\sqrt{\frac{2}{\pi}}\frac{\overline{\rho}}{M}
\frac{\delta_c}{\sigma^2(R,\tau)}\frac{d\sigma(R,\tau)}{dM}
e^{-\frac{\delta_c^2}{2\sigma^2(R,\tau)}}dM.
}
The number of objects with masses greater than a given $M$ equals 
$$
N(>M)=\int_M^{\infty} n(M^\prime)dM^\prime.
$$
Dependence of the mass function on cosmological parameters is shown in 
Fig.~\ref{multy_mf}a,b. For numerical calculations, we have 
used an analytic form of  power spectrum  ~\cite{dan}  normalized 
according to ~\cite{bunn}. 
Filled circles depict experimental data taken from 
~\cite{gir}. 
Long dashed lines depict the mass functions which differ from 
the previous ones by using canonical value $\delta_c=1.686$ 
instead of their exact values   
$\delta_c\equiv\delta_c(\Omega_m,\Omega_{\Lambda})$. 
As we can see in Fig.~\ref{multy_mf}a-c the difference in the logarithmic scale is 
small and lesser than experimental errors. 
While switching to linear scales they appear to be more prominent. 
Fig.~\ref{Multy_mfer} represents graphs for relative errors of the mass function 
for different cosmological models with fixed values of $\Omega_{\Lambda}$ (a), 
and for models with zero curvature (b)
$$Error=\frac{N(>M,\delta_c(\Omega_m,\Omega_{\Lambda}))-N(>M,\delta_c=1.686)}
{N(>M,\delta_c(\Omega_m,\Omega_{\Lambda}))}\times 100\%.$$
As we can see, the relative difference of the concentrations of rich clusters 
of galaxies, calculated according to the Press-Schechter formalism with real and 
canonical values $\delta_c$, is quite large  
and lie within $10 \div 20\%$. For the closest zero-curvature model with  
$\Omega_m=0.36$ аnd $\Omega_{\Lambda}=0.64$ (see Fig.~\ref{multy_mf}b), the relative 
error changes in the range of 5 to 20 percents.  
This model is also superior compared to other tests ~\cite{dn01,dn02}.
For some other models, this error can be no less than several dozen percents 
(Fig.~\ref{Multy_mfer}).
In fact, as can be deduced from (\ref{w2}), 
$$
\frac{\Delta n}{n}=\left|1-\frac{\delta_c^2}{\sigma^2}\right| 
\left|\frac{\Delta \delta_c}{\delta_c}\right|,
$$
where $\Delta \delta_c$ is difference between model $\delta_c$ and its canonical
value 1.686. For $\sigma(M) \le 0.8$ and 
$|\Delta \delta_c / \delta_c| \sim 0.06$
$\Delta n/n$ (and correspondingly $Error$) will be $> 20\%$.
This fact suggests that one should take into account dependence of critical 
amplitude on cosmological parameters, in order to keep calculation errors 
within 10\%.
The observational mass function data do not prevent one from possible realization 
of the model with a curvature other than zero. For example, Fig.~\ref{Multy_mfer} 
demonstrates the most consistent with experimental data mass function for the model 
with a negative curvature $\Omega_k=0.1$. Theoretical mass 
functions in the case of elliptical collapse Sheth-Tormen (ST) \cite{ST} and 
Lee-Shandarin (LS) \cite{LS} are presented here also. 
As we can see, the theory of spherical collapse well agrees with 
experimental data, since rich clusters of galaxies have formed in maxima of the 
density peaks which predominantly have spherically symmetrical form.

\section{CONCLUSIONS}

We have analyzed the evolution of a spherical dust-like cloud
from the early linear stage, when the relative difference between the background 
and densities of cloud is $\delta \ll 1$, up to the late non-linear stage, when 
the collapse occurs, $\delta\to\infty$, in cosmological models with an 
arbitrary 3D space curvature and positive values of cosmological constant. 
There are two parameters convenient for comparison of such dynamics in various
cosmological models.
First, the threshold value of initial amplitude of density perturbation, $\delta_c$, 
calculated at the moment of collapse according
to the law of linear growth. The second is the final amplitude of overdensity following
the dynamical equilibrium, $\Delta_c$. $\delta_c$ and $\Delta_c$ are used 
for calculation of the mass function and X-ray temperature of rich clusters
of galaxies respectively.
On the base of the analytical and numerical results presented here it 
can be concluded that for models with negative or zero 3D space curvature and 
positive cosmological constant ($\Lambda>0$), $\delta_c$ is smaller than its 
canonical value 1.686 ($k=\Lambda=0$)
due to longer duration of perturbation increasing. It decreases with increase of
$\Omega_m$ and with decrease of $\Omega_k$. 

In models with positive 3D 
curvature $\delta_c$ can be either greater or smaller in comparison with canonical
value. In this case critical amplitude
$\delta_c$ goes to infinity when $\Omega_m$ converges to some
small critical value.
It happens because the moment of slow down of expansion of the background
takes place.
At this moment amplitude of density perturbations grows very quickly tending 
to the exponential law in models with collapse. Such models are gravitationally 
unstable in the sense that even the smallest perturbations cause their 
collapse in finite time. 

The range of change for the threshold value  
$\delta_c$, within the range of parameters $-0.4\le\Omega_k\le0.4$, 
$0\le\Omega_{\Lambda}\le 1$, $0.1\le\Omega_m\le 1$ is small:  
$1.55\le \delta_c\le 1.75$. But the discrepancy in values of the concentration of 
rich clusters of galaxies, as calculated according to the Press-Schechter formalism, 
with real and canonical values $\delta_c$, reaches few dozens of percents for 
some models. This discrepancy grows with an increase of the mass of clusters. 

The range of variation for $\Delta_c$ is much wider,  $60\le\Delta_c\le180$ 
(for the same range of cosmological parameters $\Omega_k$, $\Omega_{\Lambda}$ and 
$\Omega_m$). This leads 
to the deviations of the calculated values of the X-ray temperature of the gas by 
$\sim40\%$, as compared to those calculated for the canonical value of $\Delta_c$, $178$.

Experimental uncertainties of determining  mass function and X-ray 
temperature function of rich clusters of galaxies in the vicinity of Milky Way are 
considerably larger than those for the models with $\Omega_m\sim 0.4$, 
$\Omega_{\Lambda}\sim 0.6$ and $\Omega_k\sim 0$ 
(see \cite{dn01} and references therein). Nevertheless, 
with accumulation of data using deep-sky 
surveys (including SDSS, 2dF GRS etc) and with respective increase of accuracy 
of the data observed,  the given dependences of $\delta_c$ and $\Delta_c$ on 
$\Omega_m$ and $\Omega_{\Lambda}$ should be taken into account when comparing 
theoretical predictions with the observational data in problems aimed searching for 
adequate models of the Universe.

\section{Acknowledgments}

Authors are thankful to Dr. S. Apunevych and Dr. Yu. Chornij for useful discussions
and comments. BN is grateful also to Astronomical observatory of Jagellonian University
in Cracow for hospitality as well as prof. M. Ostrowski and Dr. Hab. A. Wozchyna for 
discussions.


\begin{thebibliography}{}  

\bibitem{ps} W.H. Press, P. Schechter. 1974, ApJ, 187, 425
\bibitem{cav} A. Cavaliere, S. Colafranchesco, R. Scaramella. 1991, ApJ, 380, 15
\bibitem{blanch} A. Blanchard, D. Valls-Gabaud, G.A. Mamon. 1992, Astron. and Astrophys.,
 264, 365
\bibitem{eke} V.R. Eke, S. Cole and C.S. Frenk. 1993, MNRAS, 282, 266-280
\bibitem{mon} P. Monaco. 1997, MNRAS, 282, 1096; ibid, 290, 439.
\bibitem{audit} E. Audit, R. Teyssier, J.L. Alimi. 1997, Astron. and Astrophys., 325, 439
\bibitem{lee} J. Lee, S.F. Shandarin. 1998, ApJ, 500, 14
\bibitem{sheth} R. Sheth, J.H. Mo, Tormen, G. 1999, MNRAS, 307, 203
\bibitem{elena} E. Pierpaoli, D. Scott and M. White. 2001, MNRAS, 325, 77
\bibitem{rie98} A. Riess. et al. 1998, ApJ, 116, 1009
\bibitem{per98} S. Perlmutter et al. 1998, Nature, 391, 51
\bibitem{per99} S. Perlmutter et al. 1999, ApJ, 517, 565
\bibitem{lokas} E.L. Lokas, Y.Hoffman. astro-ph/0108283
\bibitem{tolm} R.C. Tolman. 1969, Relativity thermodynamics and cosmology. Oxford, Clarendon press
\bibitem{land} L.D. Landau, E.M. Lifshits. 1973, 'Teoriya polya'. Moskow, "Nauka"
\bibitem{core} J.A.Peacock and S.J.Dodds. 1994, MNRAS, 267, 1024
\bibitem{pibls} P.J.E. Peebles. 1980, The large-scale structure of the Universe. 
Princeton Univercity Press, Princeton, New Jersey
\bibitem{om1} B.S. Novosyadlyj, Yu.B. Chornyj. 1993, Astronomycheskij Zhurnal, 70, 657
\bibitem{lahav} O. Lahav, P.B. Lilje, J.R. Primack and M.J. Rees. 1991, MNRAS, 251, 128 
\bibitem{sim} D.M. White. 1993, Formation and Evolution of Galaxies.
Lectures given at Les Hauches
\bibitem{shand} N. Rahman and S.F. Shandarin. 2001, ApJ, 550L, 121R
\bibitem{dan} D.J. Eisenstein and W. Hu. 1999, ApJ, 511, 5E
\bibitem{bunn} E.F. Bunn, M. White. 1997, ApJ, 480, 6B
\bibitem{dn01} R. Durrer and B. Novosyadlyj. 2001, MNRAS, 324, 560 
\bibitem{dn02} R. Durrer, B. Novosyadlyj, S. Apunevych. 2003, ApJ, 583, 33 (astro-ph/0111594)
\bibitem{ST} R.K. Sheth, G. Tormen. 1999, MNRAS, 308, 119
\bibitem{LS} J. Lee, S.F. Shandarin. 1999, ApJ, 517, L5
\bibitem{gir} M. Girardi, S. Borgani, G. Giuricin, F.Mardirossian, M. Mazzetti. 
1998, ApJ, 506, 45G
\end{thebibliography}
\end{document}